\documentstyle[12pt]{article}
\def\doublespace{\def\baselinestretch{1.3}\Large\normalsize}
\title{\bf Supersymmetry and the Hartmann Potential of Theoretical Chemistry
\thanks{to be published at the  {\it Theoretica Chimica Acta
}}}
\author{ GARDO GARNET BLADO \thanks{e-mail: bladogg@cda.mrs.umn.edu; FAX: (612) 589-6371}\\
         \it Division of Science and Mathematics
         University of Minnesota, Morris\\
            \it Morris, MN, U.S.A. 56267-2128}
\date{}
\begin{document}
\doublespace
\maketitle
\begin{abstract}
The ring-shaped Hartmann potential
$
V = \eta \sigma^{2} \epsilon_{0} \left( \frac{2 a_{0}}{r} - \frac{\eta a_{0}^{2}}{r^{2} sin^{2} \theta}    \right)
$
was introduced in quantum chemistry to describe ring-shaped molecules like benzene. In this article, the supersymmetric features
of the Hartmann potential are discussed.
We first review
the results of a previous paper in which we rederived the eigenvalues and radial eigenfunctions of the Hartmann
potential using a formulation of one-dimensional supersymmetric quantum mechanics (SUSYQM) on the half-line
$\left[ 0, \infty \right)$. A reformulation of SUSYQM in the full line $\left( -\infty, \infty \right)$ is subsequently
developed. It is found that the second formulation makes a connection between states having the same quantum number
$L$ but different values of 
$\eta \sigma^{2}$ and quantum number $N$. This is in contrast to the first formulation, which relates states with identical values of  
the quantum number $N$ and $\eta \sigma^{2}$ but different values of the quantum number $L$.
\vspace{1.0in}
\begin{tabbing}
           \> ring-shaped potential, superpotential
\end{tabbing}
\end{abstract}
\vspace{1.0in}

\clearpage

\section{Introduction}
\label{intro}
\indent

In 1972, an exactly solvable ring-shaped potential was introduced by H. Hartmann \cite{hart}. The Hartmann potential is given by
the following expression,
\begin{equation}
V = \eta \sigma^{2} \epsilon_{0} \left( \frac{2 a_{0}}{r} - \frac{\eta a_{0}^{2}}{r^{2} sin^{2} \theta}    \right)
\label{1susyhh1}
\end{equation}
where
\begin{equation}
 a_{0} = \frac{\hbar^{2}}{\mu e^{2}}\;\;\;\;\;\;\;\; \rm and \;\;\;\;\;\;\;\; \mit \epsilon_{0} = -\frac{1}{2} \frac{\mu e^{4}}{\hbar^{2}}
\end{equation}
$\mu$ is the particle mass, $\eta$ and $\sigma$ are positive real parameters which range from about 1 to 10 in theoretical chemistry applications
\cite{schuch} and $r$, $\theta$ are in spherical coordinates. In a previous paper \cite{blado}, SUSYQM techniques were used to rederive
the eigenvalues and eigenfunctions of the Hartmann potential. In this article, we further explore its supersymmetric quantum mechanical 
features by a second formulation of SUSYQM. This analysis is inspired by the analysis made for the hydrogen atom 
using supersymmetry (SUSY) in reference \cite{hay}.

The concept of supersymmetry (SUSY) has been used in particle physics in the past two decades \cite{nil,haber}. It was discovered in 1971
by Gel'fand and Likhtman \cite{gelfand}. Simply put, supersymmetry is a symmetry which relates fermionic and bosonic degrees
of freedom. At present, particle physicists believe that it is an essential ingredient in unifying the four fundamental forces
in nature namely the electromagnetic, weak, strong and gravitational interactions.

Supersymmetric theories of the four fundamental interactions entail the presence of SUSY partners which have the same mass as their
corresponding ordinarily observed particles.\footnote{For example, the SUSY partner of the electron is called a selectron while that of
a photon is a photino.}
Unfortunately, these have not
been observed in nature. To make sense out of this experimental fact, theorists believe that SUSY must be ``broken" at ordinary
energies. The search for a mechanism to break SUSY led Witten \cite{witten} in 1981 to study SUSY breaking in the simplest
case of SUSY quantum mechanics. In fact, studies in SUSYQM during its early years, were confined solely for understanding SUSY 
breaking.

However, it was eventually discovered that SUSYQM can have interesting applications besides its use in the study of SUSY
breaking. At present, it has found its way in many areas of physics including atomic physics, statistical physics, nuclear
physics. etc. \cite{cooper} and most recently in quantum chemistry \cite{blado}. 
Through the present article, the author hopes to contribute towards the further utilization of SUSYQM in 
theoretical chemistry.

In section \ref{susyqm}, a pedagogical introduction to SUSYQM is developed. Only the concepts and equations 
which are essential 
for the present paper are presented.

We present the the SUSYQM features  of the Hartmann potential in section \ref{susyhh}. In subsection \ref{eigen}, we review our 
results from reference \cite{blado} \footnote{We do this for completeness and to facilitate the comparison with the second formulation.}
where we formulated SUSYQM in the half-line $\left[ 0, \infty \right)$. Subsection \ref{eta} discusses 
the second formulation of SUSYQM in the full-line 
$\left( -\infty, \infty \right)$. Subsequently, this formulation is  compared with the first formulation of subsection \ref{eigen}.

We give some conclusions in section \ref{concl}.
\clearpage

\section{Supersymmetric Quantum Mechanics [3]} 
\label{susyqm}
\indent

As mentioned in the Introduction, SUSY was first applied to particle physics, whose language is quantum field theory. In quantum field theory,
a particle is represented by a component field $\varphi_{i}$ and its dynamics is described by a lagrangian density 
$\cal L \mit (\varphi_{i}, \partial_{\mu} \varphi_{i})$ where $\left[  \partial_{\mu} \equiv \left( \frac{1}{c} \frac{\partial}{\partial t}, 
\vec{\nabla}\right) \right]$. The word ``supersymmetry" was originally used to describe the symmetry which transforms a field $\varphi$ to another
field $\psi$ whose intrinsic spin differs from $\varphi$ by $\frac{1}{2} \hbar$. In SUSYQM, which will be described here, we will use the term 
``supersymmetry" in a more general sense. It will be used to denote systems which can be described by the SUSY algebra.

SUSYQM \cite{witten,suku} is characterized by the existence of the charge operators $Q_{i}$, where $i = 1, 2, \ldots, N$ such that they obey the SUSY algebra
(denoted by sqm($N$)), 
\begin{equation}
\left\{  Q_{i}, Q_{j} \right\} = \delta_{ij} H_{ss}\;\;\;\;\;\;\;\; \left[  Q_{i}, H_{ss} \right] = 0
\label{1susyqm1}
\end{equation}
where $H_{ss}$ is the supersymmetric Hamiltonian, \{ \} and [ ] are anticommutator and commutator respectively. We consider only sqm(2)
with charge operators $Q_{1}$ and $Q_{2}$ and construct the linear combinations
\begin{equation}
Q = \frac{1}{\sqrt{2}} \left( Q_{1} + i Q_{2}  \right) \;\;\;\;\;\;\;\; \rm and \;\;\;\;\;\;\;\; \mit Q^{\dagger} = 
\frac{1}{\sqrt{2}} \left( Q_{1} - i Q_{2}  \right)\;.
\label{2susyqm1}
\end{equation}
From equations \ref{1susyqm1}  and \ref{2susyqm1}, The SUSY algebra is then 
\begin{equation}
\left\{  Q, Q^{\dagger} \right\} = H_{ss}, \;\;\;\;\;\; Q^{2} = 0, \;\;\;\;\;\; \left( Q^{\dagger}  \right)^{2} = 0
\label{3susyqm1}
\end{equation}
with
\begin{equation}
\left[  Q, H_{ss} \right] = 0 \;\;\;\;\;\;\;\; \rm and \;\;\;\;\;\;\;\; \mit \left[  Q^{\dagger}, H_{ss} \right] = \rm 0.
\end{equation}

The above SUSY algebra can be realized by letting 
\begin{equation}
Q = \left[ \begin{array}{cc}
0 &  0    \\
 A^{-} & 0     
\end{array} \right]
\;\;;\;\;
Q^{\dagger} = \left[ \begin{array}{cc}
0 & A^{+}     \\
0 & 0     
\end{array} \right]
\label{2susyqm2}
\end{equation}
where
\begin{equation}
\left(  A^{-} \right)^{\dagger} = A^{+}.
\end{equation}
From equations \ref{3susyqm1} and \ref{2susyqm2}, the supersymmetric hamiltonian is
\begin{equation}
H_{ss} = \left[ \begin{array}{cc}
H_{1} &  0    \\
 0 & H_{2}     
\end{array} \right]
\label{3susyqm2}
\end{equation}
where
\begin{equation}
H_{1} = A^{+} A^{-}\;\;\;\;\;\;\;\; \rm and \;\;\;\;\;\;\;\; \mit H_{2} = A^{-} A^{+}.
\label{4susyqm2}
\end{equation}
The hamiltonians $H_{1}$ and $H_{2}$ are said to be ``supersymmetric" partners of each other.
$H_{1}$ is called the "Bose" sector while $H_{2}$ is the "Fermi" sector.

The hamiltonian of the Schr\"{o}dinger equation can always be factorized in the form of equation \ref{4susyqm2}. Consider the hamiltonian
\cite{cooper}
\begin{equation}
H_{1} = -\frac{1}{2} \frac{d^{2}}{dx^{2}} + V_{1}(x)
\label{5susyqm2}
\end{equation}
such that
\begin{equation}
H_{1} \psi^{n}_{(1)} = \left[ -\frac{1}{2} \frac{d^{2}}{dx^{2}} + V_{1}(x)  \right] \psi^{n}_{(1)} = E^{n}_{(1)} \psi^{n}_{(1)}
\label{5psusyqm2}
\end{equation}
where $V_{1}(x)$ is chosen such that the ground state $\psi^{0}_{(1)}$ has an energy eigenvalue equal to zero. The hamiltonian in equation 
\ref{5susyqm2} can be put in the form of equation \ref{4susyqm2} by letting
\begin{equation}
A^{-}_{1} = \frac{1}{\sqrt{2}} \left( \frac{d}{dx} + W_{1}  \right)\;\;\;\;\;\;\;\; \rm and \;\;\;\;\;\;\;\; \mit 
A^{+}_{\rm 1} = \frac{\rm 1}{\sqrt{\rm 2}} \left( -\frac{d}{dx} + W_{\rm 1}  \right)
\label{6susyqm2}
\end{equation}
provided that the ``superpotential'' $W_{1}$ satisfies the Ricatti equation
\begin{equation}
V_{1}(x) = \frac{1}{2} \left[  W_{1}^{2} - \frac{dW_{1}}{dx}\right]\;.
\label{1susyqm3}
\end{equation}
As long as equation \ref{1susyqm3} has a solution $W_{1}$, the one-dimensional Schr\"{o}dinger equation can be made supersymmetric
by the construction given in equations \ref{6susyqm2}, \ref{4susyqm2} and \ref{3susyqm2}. The challenge then in using SUSYQM techniques
is not in the mechanics of the construction of just any SUSY hamiltonian, but in finding a suitable superpotential (or $V_{1}(x)$) to 
construct a SUSY hamiltonian which will be relevant to the problem at hand. It is a common practice to choose or pose as an ansatz the 
$W_{1}$ to solve a physical problem \cite{schwabl,rico}.

The SUSY partner of $H_{1}$, namely $H_{2}$ is then given by
\begin{equation}
H_{2} = -\frac{1}{2} \frac{d^{2}}{dx^{2}} + V_{2}(x) = A^{-}_{1} A^{+}_{1}
\label{2susyqm3}
\end{equation}
where
\begin{equation}
V_{2}(x) = \frac{1}{2} \left[  W_{1}^{2} + \frac{dW_{1}}{dx}\right]\;.
\label{5-16}
\end{equation}

Note that $H_{2}$ is altogether a new hamiltonian. An astute reader will immediately realize that one can repeat the procedure in constructing $H_{2}$ from $H_{1}$ to construct an $H_{3}$ from $H_{2}$ such that 
\begin{equation}
H_{2} = A^{+}_{2} A^{-}_{2}
\end{equation}
with
\begin{equation}
A^{-}_{2} = \frac{1}{\sqrt{2}} \left( \frac{d}{dx} + W_{2}  \right)\;\;\;\;\;\;\;\; \rm and \;\;\;\;\;\;\;\; \mit 
A^{+}_{\rm 2} = \frac{\rm 1}{\sqrt{\rm 2}} \left( -\frac{d}{dx} + W_{\rm 2}  \right)
\end{equation}
and with a new Ricatti equation
\begin{equation}
V_{2}(x) = \frac{1}{2} \left[  W_{2}^{2} - \frac{dW_{2}}{dx}\right].
\label{1susyqm4}
\end{equation}
$W_{2}$ in equation \ref{1susyqm4} is then solved to construct $A^{\pm}_{2}$. $H_{3}$ can then be constructed as
\begin{equation}
H_{3} = -\frac{1}{2} \frac{d^{2}}{dx^{2}} + V_{3}(x) = A^{-}_{2} A^{+}_{2}
\end{equation}
where
\begin{equation}
V_{3}(x) = \frac{1}{2} \left[  W_{2}^{2} + \frac{dW_{2}}{dx}\right]\;.
\end{equation}
We can evidently construct a ``hierarchy" of SUSY-partner hamiltonians $H_{1}$, $H_{2}$, $H_{3}$, \ldots , $H_{n}$ starting from $H_{1}$.

Let us go back to the first two hamiltonians we started with namely $H_{1}$ and $H_{2}$. Since $V_{1}(x)$ in equation \ref{5susyqm2} is chosen
such that its ground state wave function $\psi^{0}_{(1)}$ has eigenvalue equal to zero, equation \ref{4susyqm2} gives 
\begin{equation}
H_{1} \psi^{0}_{(1)} = 0 \;\;\; \Longrightarrow \;\;\; A^{+}_{1} A^{-}_{1} \psi^{0}_{(1)} = 0.
\label{4susyqm4}
\end{equation}
Equation \ref{4susyqm4} implies
\begin{equation}
A^{-}_{1} \psi^{0}_{(1)} = 0.
\label{5susyqm4}
\end{equation}
With equation \ref{6susyqm2} and knowing $W_{1}$, equation \ref{5susyqm4} allows one to calculate the ground state of $H_{1}$ by solving the
resulting first order differential equation. 

Consider any eigenstate of $H_{1}$, $\psi^{n}_{(1)}$ with energy $E^{n}_{(1)}$. We have
\begin{equation}
H_{1} \psi^{n}_{(1)} = E^{n}_{(1)}\psi^{n}_{(1)}
\end{equation}
or from equation \ref{4susyqm2} 
\begin{equation}
A^{+}_{1} A^{-}_{1} \psi^{n}_{(1)} = E^{n}_{(1)} \psi^{n}_{(1)}.
\label{1susyqm5}
\end{equation}
Applying $A^{-}_{1}$ to equation \ref{1susyqm5},
$A^{-}_{1} A^{+}_{1} \left( A^{-}_{1} \psi^{n}_{(1)}  \right) = E^{n}_{(1)} A^{-}_{1} \psi^{n}_{(1)}$ 
or with equation \ref{2susyqm3} 
\begin{equation}
H_{2} \left( A^{-}_{1} \psi^{n}_{(1)}  \right) = E^{n}_{(1)} \left( A^{-}_{1} \psi^{n}_{(1)} \right).
\label{2susyqm5}
\end{equation}

Conversely, consider an eigenfunction $\psi^{n}_{(2)}$ of $H_{2}$ with eigenvalue $E^{n}_{(2)}$. With equation \ref{2susyqm3}, we get 
$H_{2} \psi^{n}_{(2)} = A^{-}_{1} A^{+}_{1} \psi^{n}_{(2)} = E^{n}_{(2)} \psi^{n}_{(2)}$. Multiplying by $A^{+}_{1}$, we have,
$A^{+}_{1} A^{-}_{1} \left( A^{+}_{1} \psi^{n}_{(2)}  \right) = E^{n}_{(2)} A^{+}_{1} \psi^{n}_{(2)}$. With equation \ref{4susyqm2}
\begin{equation}
H_{1} \left( A^{+}_{1} \psi^{n}_{(2)}  \right) = E^{n}_{(2)} \left( A^{+}_{1} \psi^{n}_{(2)} \right).
\label{3susyqm5}
\end{equation}

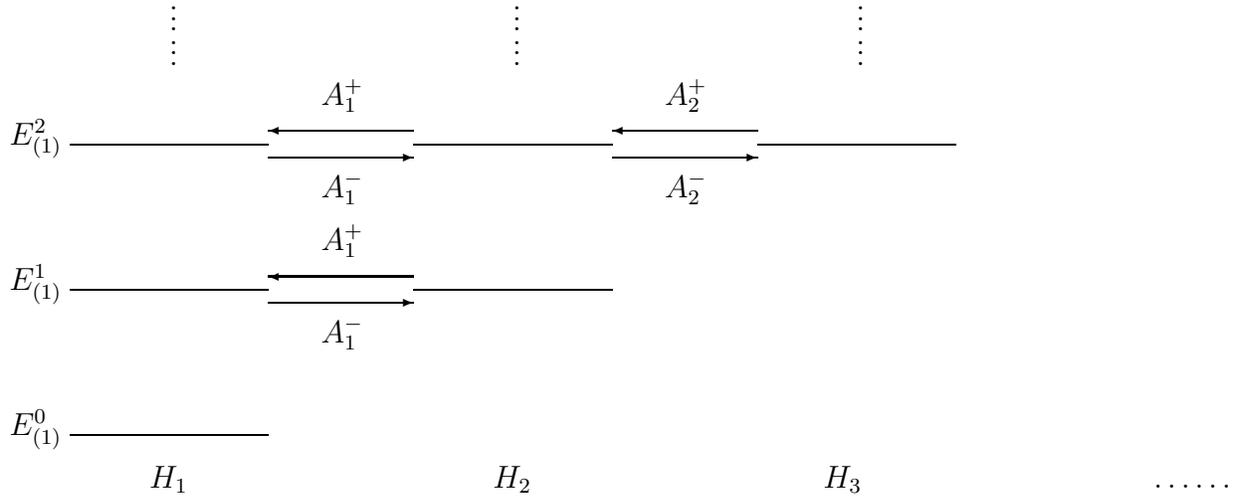
\begin{figure}[htbp]
\begin{center}
\begin{picture}(400,200)
%%%%%%%%%%%%%%%%%%%%%%%%%%%%%
                            \put(37.5,173){\vdots}                                      \put(167.5,173){\vdots}                                       \put(297.5,173){\vdots}
                            \put(37.5,160){\vdots}                                      \put(167.5,160){\vdots}                                       \put(297.5,160){\vdots}
                                                            \put(95,145){$A^{+}_{1}$}                                     \put(225,145){$A^{+}_{2}$}
                                                            \put(130,135){\vector(-1,0){55}}                              \put(260,135){\vector(-1,0){55}}
\put(-23,130){$E^{2}_{(1)}$} \put(0,130){\line(75,0){75}}                                \put(130,130){\line(75,0){75}}                                \put(260,130){\line(75,0){75}}
                                                            \put(75,125){\vector(1,0){55}}                                \put(205,125){\vector(1,0){55}}
                                                            \put(95,110){$A^{-}_{1}$}                                     \put(225,110){$A^{-}_{2}$}
                                                            \put(95,90){$A^{+}_{1}$}
                                                            \put(130,80){\vector(-1,0){55}}
\put(-23,75){$E^{1}_{(1)}$} \put(0,75){\line(75,0){75}}                                 \put (130,75){\line(75,0){75}}
                                                            \put(75,70){\vector(1,0){55}}
                                                            \put(95,55){$A^{-}_{1}$}
\put(-23,20){$E^{0}_{(1)}$} \put(0,20){\line(75,0){75}}
                            \put(30,0){$H_{1}$}                                         \put(160,0){$H_{2}$}                                           \put(285,0){$H_{3}$}           \put(410,0){\ldots \ldots}
%%%%%%%%%%%%%%%%%%%%%%%%%%%%
\end{picture}
\end{center}
\caption{The hierarchy of hamiltonians and the action of the operators $A^{\pm}_{n}$ on the degenerate eigenstates}
\label{fig1susyqm5}
\end{figure}

Equations \ref{2susyqm5} and \ref{3susyqm5} imply that the hamiltonians $H_{1}$ and $H_{2}$ have identical eigenvalues except for the ground state
$\psi^{0}_{(1)}$ of $H_{1}$ (since $A^{-}_{1} \psi^{0}_{(1)}=0$ in equation \ref{2susyqm5} and this is unnormalizable). In addition, we  can see that
if you know an eigenfunction of $H_{1}$, i.e. $\psi^{n}_{(1)}$, then an eigenfunction $A^{-}_{1} \psi^{n}_{(1)}$ of $H_{2}$ can be formed. Similarly,
an eigenfunction $A^{+}_{1} \psi^{n}_{(2)}$ of $H_{1}$ can be formed given an eigenfunction $\psi^{n}_{(2)}$ of $H_{2}$. The preceding analysis can
then be extended to the hierarchy of hamiltonians discussed earlier. These observations are illustrated in figure \ref{fig1susyqm5}.

Herein lies a very important consequence of SUSYQM. The energy eigenfunctions of the hierarchy of hamiltonians are related by the 
SUSY operators $A^{\pm}$. If one knows the eigenvalues and eigenfunctions of a particular $H_{n}$, then  one can get the eigenvalues and eigenfunctions of its SUSY
partner.

Note that what we have discussed is SUSYQM in one dimension.There had been work done in doing SUSYQM in two or more dimensions \cite{filho,andria}. To be able to apply
one dimensional SUSYQM to the Hartmann potential, we will do a separation of variables of the resulting Schr\"{o}dinger equation. In forming  a hamiltonian
from
 a separated one dimensional
differential equation that can be an element of an $H_{ss}$, this one dimensional differential equation
 must be of the form of equation \ref{5psusyqm2} (no first derivative term) and must yield an infinite tower of 
states as for $H_{n}$ of figure \ref{fig1susyqm5} \cite{bluhm}. As we will see, upon separation of variables for the Hartmann potential, only the radial equation yields
an interesting SUSY.

\clearpage

\section{Supersymmetry in the Hartmann Hamiltonian}
\label{susyhh}
\indent

With the SUSYQM concepts introduced in section \ref{susyqm}, we are now ready to discuss the SUSY features of the radial equation of the
Schr\"odinger equation of the Hartmann potential in spherical coordinates. 

\subsection{Calculation of the eigenvalues and radial eigenfunctions}
\label{eigen}
\indent

The Schr\"{o}dinger equation in spherical coordinates for a particle of mass $\mu$ subjected to the Hartmann potential in equation \ref{1susyhh1}
is given by
\begin{equation}
-\frac{\hbar^{2}}{2 \mu} \nabla^{2} \psi + \left[ \frac{2 \eta \sigma^{2} \epsilon_{0} a_{0}}{r}  - 
\frac{\eta^{2} \sigma^{2} a_{0}^{2} \epsilon_{0}}{r^{2} sin^{2} \theta}   \right] \psi = E \psi\;.
\label{3susyhh1}
\end{equation}
Assuming a solution
\begin{equation}
\psi = R(r) \Theta(\theta) \Phi(\phi)
\end{equation}
equation \ref{3susyhh1} can be separated into three differential equations \cite{hart} 
\begin{equation}
\frac{1}{\Phi} \frac{d^{2} \Phi}{d \phi^{2}} = - m^{2}
\label{1susyhh2}
\end{equation}
\begin{equation}
\frac{1}{sin \theta} \frac{d}{d \theta} \left( sin \theta \frac{d \Theta}{d \theta} \right) - 
\left( \frac{M^2}{sin^2 \theta} - L(L+1) \right) \Theta = 0
\label{2susyhh2}
\end{equation}
\begin{equation}
\frac{1}{r^2} \frac{d}{dr} \left( r^2 \frac{dR}{dr} \right) - L(L+1) \frac{R}{r^2} +
\frac{8 \pi^2 \mu}{h^2} \left( E + \frac{\eta \sigma^2 e^2}{r} \right) R = 0
\label{3susyhh2}
\end{equation}
where
\begin{equation}
M^2 = m^2 + \eta^2 \sigma^2 \;.
\end{equation}

Looking at equations \ref{1susyhh2} to \ref{3susyhh2}, we realize that these closely resemble the separated equations of the hydrogen atom \cite{grif}.
As shown by reference \cite{bluhm}, the only interesting separable SUSY in the hydrogen atom in spherical coordinates results from the radial equation.
Their argument is as follows. Looking at equation \ref{1susyhh2}, and comparing it with equation \ref{5psusyqm2}, we see that $V_{1} = 0$. Equation 
\ref{2susyhh2} on the other hand can be cast to a form similar to equation \ref{5psusyqm2} by multiplying it by a modulation factor 
$\frac{f(cos \theta)}{\left[ 1 - cos^2 \theta  \right]^{1/2}}$. The eigenvalue of $f(cos \theta)$  is zero and no infinite tower of states can be generated. Hence,
the $\Phi$ and $\Theta$ solutions cannot be given by SUSYQM. They can be solved by conventional means \cite{hart}
\begin{equation}
\Phi(\phi) = \frac{1}{\sqrt{2 \pi}} e^{i m \phi}, \;\;\;\;\; m = 0, \;\pm 1,\; \pm 2, \ldots
\end{equation}
\begin{equation}
\Theta (\theta) \sim \cal P ^{\left| \mit M \right|}_{\mit L} (\mit cos \theta), \;\;\;\;\; \mit L = \nu^{\prime} + \left| \mit M \right| ,
\;\;\;\;\; \nu^{\prime} = \rm 0,\;1,\;2, \ldots
\label{6susyhh2}
\end{equation}
where $\cal P ^{\left| \mit M \right|}_{\mit L} (\mit cos \theta)$ are the associated Legendre polynomials.

The radial equation \ref{3susyhh2} can be cast into a form similar to \ref{5psusyqm2} by letting 
\begin{equation}
R = \frac{u}{r}\;.
\label{1susyhh3}
\end{equation}
Substituting equation \ref{1susyhh3} into equation \ref{3susyhh2}, yields 
\begin{equation}
H_{L} u = \left[ -\frac{1}{2} \frac{d^2}{dr^2} +  \frac{L(L+1)}{2 r^2} - \frac{\gamma}{r} \right] u = \frac{\mu E}{\hbar^2} u
\label{2susyhh3}
\end{equation}
with
\begin{equation}
\gamma \equiv \frac{\mu \eta \sigma^2 e^2}{\hbar^2}.
\label{9-38}
\end{equation}

Equation \ref{2susyhh3} is similar to that of the hydrogen atom's radial equation. We thus claim that we can obtain the eigenvalues and radial
eigenfunctions by looking at the hamiltonian \cite{schwabl}
\begin{equation}
\cal H_{\mit L} = \mit -\frac{\rm 1}{\rm 2} \frac{d^2}{dr^2} + \frac{L(L+\rm 1)}{\rm 2 \mit r^{2}} - \frac{\gamma}{r} + \frac{\rm 1}{\rm 2} \left( \frac{\gamma}{L+\rm 1}  \right)^2
\label{4susyhh3}
\end{equation}
which yields a Ricatti equation (from equations \ref{4susyhh3}, \ref{5susyqm2} and \ref{1susyqm3})
\begin{equation}
\frac{L(L+\rm 1)}{\rm 2 \mit r^{2}} - \frac{\gamma}{r} + \frac{\rm 1}{\rm 2} \left( \frac{\gamma}{L+\rm 1}  \right)^2 = 
\frac{1}{2} \left[  W_{L}^{2} - \frac{dW_{L}}{dr}\right]
\end{equation}
whose solution is
\begin{equation}
W_{L} = -\frac{L+1}{r} + \frac{\gamma}{L+1}\;.
\label{6susyhh3}
\end{equation}
Equation \ref{6susyhh3} and \ref{6susyqm2} yield 
\begin{equation}
A^{\pm}_{L} = \frac{1}{\sqrt{2}} \left( \mp \frac{d}{dr} -\frac{L+1}{r} + \frac{\gamma}{L+1} \right).
\label{1susyhh4}
\end{equation}

From equation \ref{1susyhh4} and \ref{2susyqm3}, we construct the SUSY-partner hamiltonian of $\cal H_{\mit L}$ in equation \ref{4susyhh3},
\begin{equation}
\cal H_{\mit L + \rm 1} = \mit A^{-}_{L} A^{+}_{L} =
\mit -\frac{\rm 1}{\rm 2} \frac{d^2}{dr^2} + \frac{(L + \rm 1)(\mit L+\rm 2)}{\rm 2 \mit r^{2}} - \frac{\gamma}{r} + \frac{\rm 1}{\rm 2} \left( \frac{\gamma}{L+\rm 1}  \right)^2\; .
\label{2susyhh4}
\end{equation}

Comparing equations \ref{2susyhh3} and \ref{4susyhh3} and with equation \ref{2susyhh4}, we realize that
\begin{equation}
\cal H_{\mit L} = \mit H_{L} + \frac{\rm 1}{\rm 2} \left( \frac{\gamma}{L+\rm 1}  \right)^2
\label{3susyhh4}
\end{equation}
\begin{equation}
\cal H_{\mit L + \rm 1} = \mit H_{L + \rm 1} + \frac{\rm 1}{\rm 2} \left( \frac{\gamma}{L+\rm 1}  \right)^2\; .
\label{4susyhh4}
\end{equation}

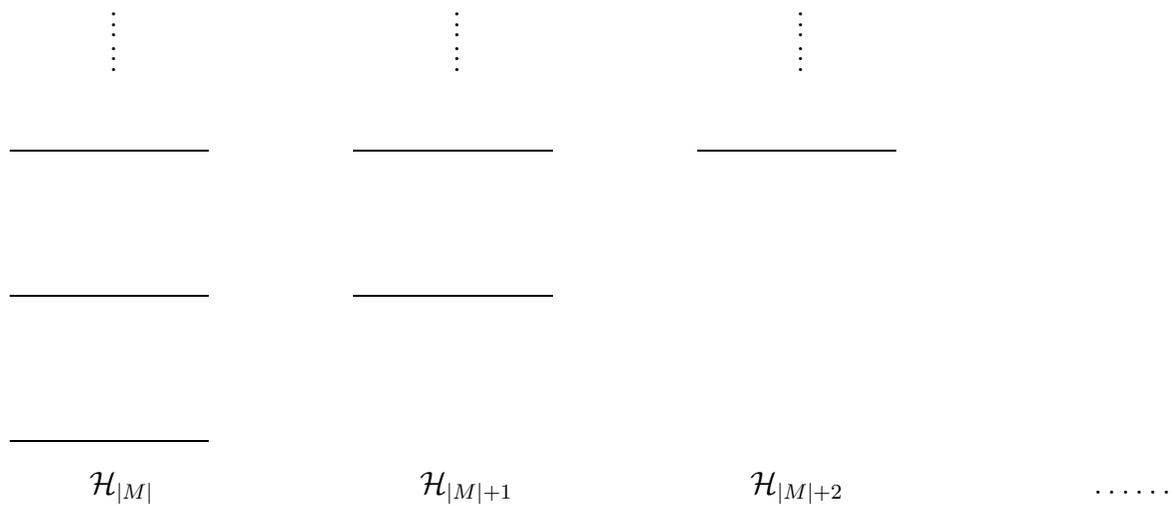
\begin{figure}[htbp]
\begin{center}
\begin{picture}(400,200)
%%%%%%%%%%%%%%%%%%%%%%%%%%%%%
                            \put(37.5,173){\vdots}                                      \put(167.5,173){\vdots}                                       \put(297.5,173){\vdots}
                            \put(37.5,160){\vdots}                                      \put(167.5,160){\vdots}                                       \put(297.5,160){\vdots}
                            \put(0,130){\line(75,0){75}}                                \put(130,130){\line(75,0){75}}                                \put(260,130){\line(75,0){75}}
                            \put(0,75){\line(75,0){75}}                                 \put (130,75){\line(75,0){75}}
                           \put(0,20){\line(75,0){75}}
                            \put(30,0){$\cal H_{\mit \left| M \right|}$}                                \put(155,0){$\cal H_{\mit \left| M \right| + \rm 1}$}                                 \put(280,0){$\cal H_{\mit \left| M \right|  + \rm 2}$}           \put(410,0){\ldots \ldots}
%%%%%%%%%%%%%%%%%%%%%%%%%%%%
\end{picture}
\end{center}
\caption{The energy states of the hierarchy of SUSY-partner hamiltonians from the Hartmann potential}
\label{fig2susyhh4}
\end{figure}

Let us now start to build up the radial eigenfunctions and in the process get the eigenvalues. Given an $\left| M \right|$ value, the lowest $L$ value
is $L = \left| M \right|$, as can be seen in equation \ref{6susyhh2}. It is apparent from equations \ref{3susyhh4} and \ref{4susyhh4} that we can build the 
states of the hierarchy of hamiltonians as in figure \ref{fig1susyqm5}. This is illustrated in figure \ref{fig2susyhh4}.

Since $\cal H_{\mit \left| M \right|}$ ($\cal H_{\mit \left| M \right| + \rm 1}$) and $H_{\left| M \right|}$ ($H_{\left| M \right|+1}$) differ only by a constant, (see equations
\ref{3susyhh4} and \ref{4susyhh4}) every eigenfunction of $\cal H_{\mit \left| M \right|}$ ($\cal H_{\mit \left| M \right| + \rm 1}$) will be an eigenfunction
of $H_{\left| M \right|}$ ($H_{\left| M \right|+1}$). Hence, all we have to do is to solve for the eigenfunctions of $\cal H_{\mit L}$. The actual energy for $H_{L}$ can be found by letting
$H_{L}$ act on the eigenfunctions of $\cal H_{\mit L}$. 

For an arbitrary $L$, equations \ref{5susyqm4} and \ref{1susyhh4} give, for the ground states of $\cal H_{\mit L}$, $\psi^{0}_{(L)}$, the first order differential
equation
\begin{equation}
\frac{1}{\sqrt{2}} \left(  \frac{d}{dr} -\frac{L+1}{r} + \frac{\gamma}{L+1} \right) \psi^{0}_{(L)} = 0 
\label{45p}
\end{equation}
which can easily be solved as
\begin{equation}
\psi^{0}_{(L)} = \cal N_{\mit L} \; \mit r^{L+ \rm 1} exp(-\kappa_{L} r)
\label{1susyhh5}
\end{equation}
where
\begin{equation}
\kappa_{L} \equiv \frac{\gamma}{L+1}\;.
\label{2susyhh5}
\end{equation}

\begin{figure}[htbp]
\begin{center}
\begin{picture}(400,200)
%%%%%%%%%%%%%%%%%%%%%%%%%%%%%
                            \put(37.5,173){\vdots}                                      \put(167.5,173){\vdots}                                       \put(297.5,173){\vdots}
                            \put(37.5,160){\vdots}                                      \put(167.5,160){\vdots}                                       \put(297.5,160){\vdots}
                                                                                                                                                      \put(285,135){$u_{\left| M \right| + 2}$}
                            \put(0,130){\line(75,0){75}}                                \put(130,130){\line(75,0){75}}                                \put(260,130){\line(75,0){75}}
                                                                                        \put(155,80){$u_{\left| M \right| + 1}$}
                            \put(0,75){\line(75,0){75}}                                 \put (130,75){\line(75,0){75}}
                            \put(27,25){$u_{\left| M \right|}$}
                            \put(0,20){\line(75,0){75}}
                            \put(30,0){$H_{\left| M \right|}$}                                \put(155,0){$H_{\left| M\right| + 1}$}                                 \put(280,0){$ H_{\left| M \right| +  2 }$}           \put(410,0){\ldots \ldots}
%%%%%%%%%%%%%%%%%%%%%%%%%%%%
\end{picture}
\end{center}
\caption{The hierarchy of hamiltonians of the Hartmann potential and their ground states. The $H_{L}$ here are the actual radial hamiltonian for a particular $L$ value.}
\label{fig3susyhh6}
\end{figure}
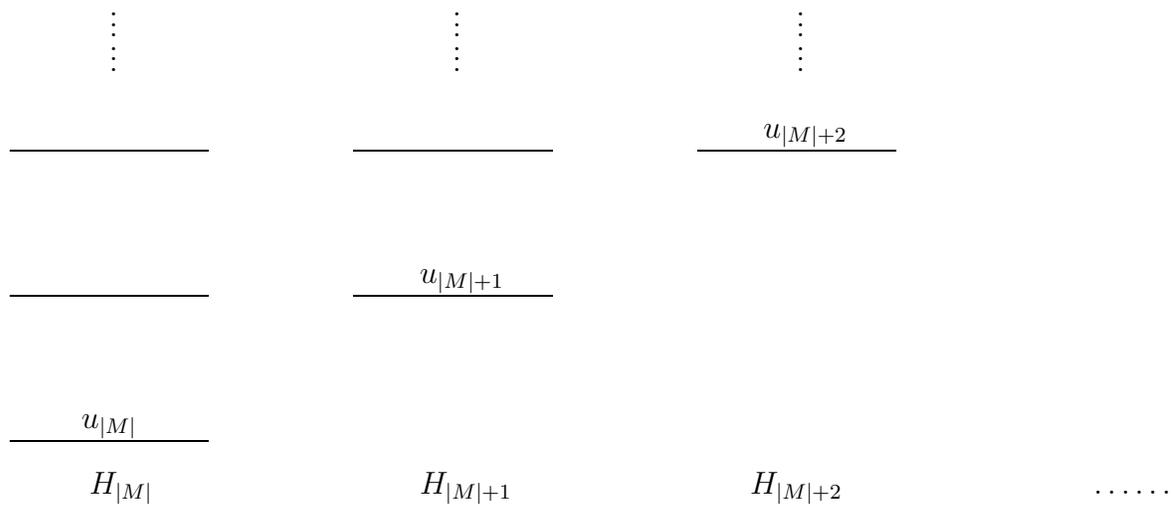

Since $L$ is arbitrary here, we realize that equation \ref{1susyhh5} is the expression for the eigenfunction for the lowest rung
(i.e. the ground state) of the tower of states for each of the 
hamiltonians in figure \ref{fig2susyhh4}. Since they are also eigenfunctions of $H_{L}$, we can write 
\begin{equation}
u_{L} = \cal N_{\mit L} \;  \mit r^{L+ \rm 1} exp(-\kappa_{L} r)
\label{3susyhh5}
\end{equation}
where we illustrate them in figure \ref{fig3susyhh6}.

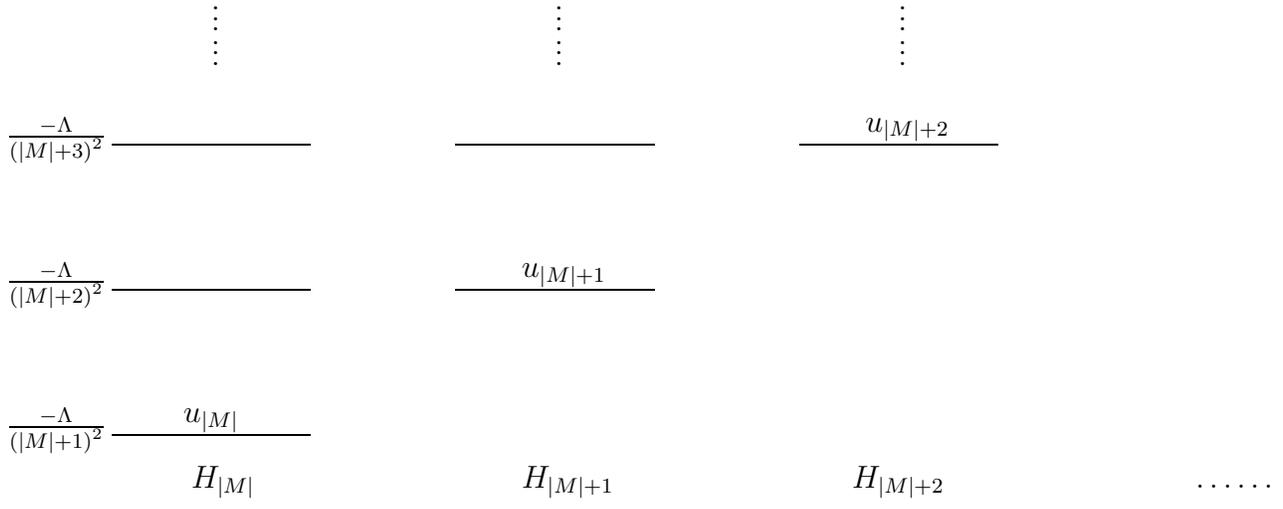
\begin{figure}[htbp]
\begin{center}
\begin{picture}(400,200)
%%%%%%%%%%%%%%%%%%%%%%%%%%%%%
                            \put(37.5,173){\vdots}                                      \put(167.5,173){\vdots}                                       \put(297.5,173){\vdots}
                            \put(37.5,160){\vdots}                                      \put(167.5,160){\vdots}                                       \put(297.5,160){\vdots}
                                                                                                                                                      \put(285,135){$u_{\left| M \right| + 2}$}
\put(-40,130){$\frac{-\Lambda}{\left( \left| M \right| +3 \right)^2}$}                            \put(0,130){\line(75,0){75}}                                \put(130,130){\line(75,0){75}}                                \put(260,130){\line(75,0){75}}
                                                                                        \put(155,80){$u_{\left| M \right| + 1}$}
\put(-40,75){$\frac{-\Lambda}{\left( \left| M \right| +2 \right)^2}$}                            \put(0,75){\line(75,0){75}}                                 \put (130,75){\line(75,0){75}}
                            \put(27,25){$u_{\left| M \right|}$}
\put(-40,20){$\frac{-\Lambda}{\left( \left| M \right| +1 \right)^2}$}                            \put(0,20){\line(75,0){75}}
                            \put(30,0){$H_{\left| M \right|}$}                                \put(155,0){$H_{\left| M\right| + 1}$}                                 \put(280,0){$ H_{\left| M \right| +  2}$}           \put(410,0){\ldots \ldots}
%%%%%%%%%%%%%%%%%%%%%%%%%%%%
\end{picture}
\end{center}
\caption{Figure 3 with the energy levels labelled.}
\label{fig4susyhh6}
\end{figure}

To get the actual energy, we let $H_{L}$ of equation \ref{2susyhh3} act on equation \ref{3susyhh5}. 
\begin{equation}
H_{L} u_{L} = \left[ -\frac{1}{2} \frac{d^2}{dr^2} + \left( \frac{L(L+1)}{2 r^2} - \frac{\gamma}{r} \right) \right] u_{L} = \frac{\mu E_{L}}{\hbar^2} u_{L}\;.
\label{10-50}
\end{equation}
After some simplification, this yields 
\begin{equation}
E_{L} = -\frac{\Lambda}{\left( L+1 \right)^2}\;\;\;\;\; \Lambda = \eta^2 \sigma^4 \left| \epsilon_{0} \right|.
\label{2susyhh6}
\end{equation}
We characterize the energy by the $L$ quantum number for the moment. From equation \ref{2susyhh6}, we can label the energy levels of figure \ref{fig3susyhh6}
as in figure \ref{fig4susyhh6}.

It is apparent from figure \ref{fig4susyhh6} that we have to label the solutions $u$ as $u_{\left| M \right| + 1,\, \left| M \right|}$; $u_{\left| M \right| + 2,\, \left| M \right| + 1}$;
$u_{\left| M \right| + 3,\, \left| M \right| + 2}$; \ldots due to the energy of the states. Knowing the eigenstates at the lowest rung of the hierarchy of
hamiltonians, $u_{\left| M \right| + 1,\, \left| M \right|}$; $u_{\left| M \right| + 2,\, \left| M \right| + 1}$;
$u_{\left| M \right| + 3,\, \left| M \right| + 2}$; etc., we can determine the other states by the action of $A^{+}_{L}$ on these eigenstates
as in figure \ref{fig1susyqm5}. This is illustrated in figure \ref{fig5susyhh7}.

Note that, for instance,  $u_{\left| M \right| + 3, \left| M \right|}$; $u_{\left| M \right| + 3, \left| M \right| + 1}$; $u_{\left| M \right| + 3, \left| M \right| + 2}$; \ldots
have the same energy $-\frac{\Lambda}{\left( \left| M  \right| + 3 \right)^2}$, and similarly for other states at the same energy level. It is then evident that given 
$N \geq \left| M \right| + 1$, $u_{N, N-1}$; $u_{N, N-2}$; $u_{N, N-3}$; \ldots; $u_{N, \left| M \right|}$ \label{page11}
will all have the same energy $-\frac{\Lambda}{N^2}$. 
Hence, we can say that 
\begin{equation}
E_{N} = -\frac{\Lambda}{N^2},\;\;\;\;\; \Lambda = \eta^2 \sigma^4 \left| \epsilon_{0} \right|,\;\;\;\;\; N = L + 1 + n^{\prime},\;\;\;\;\; n^{\prime} = 0, 1, 2, \ldots
\label{1susyhh7}
\end{equation}
which means that the energy is actually labelled by $N$ and not $L$. Equation \ref{1susyhh7} agrees with reference \cite{hart}. 

From figure \ref{fig5susyhh7}, equations \ref{2susyhh5}, \ref{3susyhh5}, \ref{1susyhh3} and \ref{1susyhh4}, it can be shown that
\begin{equation}
R_{\left| M \right| + 1, \left| M \right|}(r) = \left[ \frac{2 \gamma}{\left| M \right| + 1} \right]^{\left| M \right| + 3/2}
\left[ \frac{1}{\Gamma \left( 2 \left| M \right| + 3 \right)}  \right]^{1/2}
r^{\left| M \right|} e^{-\gamma r/\left[ \left| M \right| + 1  \right]}
\label{1susyhh8}
\end{equation}
\begin{equation}
R_{\left| M \right| + 2, \left| M \right| + 1}(r) = \left[ \frac{2 \gamma}{\left| M \right| + 2} \right]^{\left| M \right| + 5/2}
\left[ \frac{1}{\Gamma \left( 2 \left| M \right| + 5 \right)}  \right]^{1/2}
r^{\left| M \right| + 1} e^{-\gamma r/\left[ \left| M \right| + 2  \right]}
\label{2susyhh8}
\end{equation}
\begin{equation}
\begin{array}{lll}
R_{\left| M \right| + 2, \left| M \right|}(r) & = & -\left[ \frac{2 \gamma}{\left| M \right| + 2} \right]^{\left| M \right| + 3/2}
\left[ \frac{1}{2 \left( \left| M \right| + 2 \right) \Gamma \left( 2 \left| M \right| + 3 \right)}  \right]^{1/2}
r^{\left| M \right|} e^{-\gamma r/\left[ \left| M \right| + 2  \right]} \\
&  & \times \left[ 2 \left| M \right| + 2 - \frac{2 \gamma r}{ \left| M \right| + 2  } \right]
\end{array}
\label{3susyhh8}
\end{equation}
etc. where $R_{NL}$ are normalized by
\begin{equation}
1 = \int_{0}^{\infty} \left| R_{NL} \right|^2 r^2 dr
\label{4susyhh8}
\end{equation}
and the gamma function properties used are \cite{arf}
\begin{equation}
\Gamma (z) = \int_{0}^{\infty} e^{-t} t^{z-1} dt; \;\;\;\;\; \Gamma(z+1) = z \Gamma(z)\;.
\label{5susyhh8}
\end{equation}
Equations \ref{1susyhh8}, \ref{2susyhh8} and \ref{3susyhh8} agree with the normalized $R_{NL}$ of reference \cite{schuch}. It is obvious that we can eventually get all the expressions of 
$R_{NL}$.

To illustrate the procedure more concretely, let us show how we can get $R_{\left| M \right| + 2, \left| M \right|}(r)$. From figure \ref{fig5susyhh7},
\begin{equation}
u_{\left| M \right| + 2, \left| M \right|} \sim A^{+}_{\left| M \right|}\; u_{\left| M \right| + 2, \left| M \right| + 1}. 
\footnote{ Note that in equation \ref{1susyhh9} and in some of the following equations, we use $\sim$ instead of $=$ since the procedure outlined here cannot automatically normalize
the eigenfunctions.  At the end, we normalize the eigenfunctions using equation \ref{4susyhh8}. }
\label{1susyhh9}
\end{equation} 
Since $u_{\left| M \right| + 2, \left| M \right| + 1}$ is at the lowest rung of $H_{\left| M \right| + 1}$, we can use equations \ref{2susyhh5} and \ref{3susyhh5} which give
\begin{equation}
u_{\left| M \right| + 2, \left| M \right| + 1} \sim r^{\left| M \right| + 2} e^{-\gamma r/\left[ \left| M \right| + 2  \right]}\;.
\label{2susyhh9}
\end{equation}
From equations \ref{1susyhh4}, \ref{1susyhh9} and \ref{2susyhh9}, we get
\begin{equation}
u_{\left| M \right| + 2, \left| M \right|} \sim \left( - \frac{d}{dr} -\frac{\left| M \right|+1}{r} + \frac{\gamma}{\left| M \right|+1} \right) r^{\left| M \right| + 2} e^{-\gamma r/\left[ \left| M \right| + 2  \right]}
\end{equation}
which leads to 
\begin{equation}
u_{\left| M  \right| +2, \left| M  \right|} \sim r e^{-\gamma r/ \left[ \left| M  \right| + 2 \right]}
\left( -r^{\left| M  \right|} + \frac{\gamma r^{\left| M  \right| +1}}{\left(\left| M  \right| +2 \right) \left(\left| M  \right| +1 \right)}
\right).
\label{3susyhh9}
\end{equation}
Using equations \ref{3susyhh9} and \ref{1susyhh3} and rearranging terms, we get 
\begin{equation}
R_{\left| M  \right| +2, \left| M  \right|} \sim  -e^{-\gamma r/ \left[ \left| M  \right| + 2 \right]}
\;r^{\left| M  \right|} \left( 2 + 2 \left| M  \right| - \frac{2 \gamma r}{\left| M  \right| +2}
\right)
\end{equation}
or
\begin{equation}
R_{\left| M  \right| +2, \left| M  \right|} = - \cal N \mit  e^{-\gamma r/ \left[ \left| M  \right| + \rm 2 \right]}
\;r^{\left| M  \right|} \left( \rm 2 + 2 \mit \left| M  \right| - \frac{\rm 2 \mit \gamma r}{\left| M  \right| +\rm 2}
\right)
\label{5susyhh9}
\end{equation}
where $\cal N$ is the normalization constant. Normalizing equation \ref{5susyhh9} using equations \ref{4susyhh8} and \ref{5susyhh8} leads to equation \ref{3susyhh8}.

Comparing equations \ref{4susyhh3} and \ref{2susyhh4}, (discounting the common constant 
$\frac{1}{2} \left( \frac{\gamma}{L+1}   \right)^{2}$ which just rescales the ground state such that its energy eigenvalue is zero) and with figure \ref{fig5susyhh7},
we realize that states with quantum numbers ($N$, $L$) has for its SUSY partner states with quantum numbers ($N$, $L+1$).

\subsection{Relating eigenstates with different values of $\eta \sigma^2$ and quantum number N}
\label{eta}
\indent

One thing to note about the analysis of subsection \ref{eigen} is that the one dimensional SUSYQM problem was formulated in the half-line [0, $\infty$) (since 
$0\leq r< \infty$). Let us now see the consequences of formulating the SUSYQM problem in the full line ($-\infty, \infty$) \cite{hay}.

As a first step, let us rewrite 
equation \ref{2susyhh3} by a change of variables given by
\begin{equation}
y \equiv \gamma r = \frac{\left( \eta \sigma^2 \right) \mu e^2}{\hbar^2}\; r
\label{3.2-1*1}
\end{equation}
with $\gamma$ given by equation \ref{9-38}. Using equation \ref{3.2-1*1} and the equation for the energy given by equation \ref{1susyhh7}, we can rewrite equation 
\ref{10-50} as 
\begin{equation}
\left[ -\frac{1}{2} \frac{d^2}{dy^2} +  \frac{L(L+1)}{2 y^2} - \frac{1}{y}  \right] u_{L} = -\frac{1}{2 N^2} u_{N L}\;.
\label{3.2-1*2}
\end{equation}
We now make a second change of variables from $y$ to $x$ such that 
\begin{equation}
y = e^x, \;\;\;\; u_{NL} = e^{x/2} \psi
\label{3.2-1*3}
\end{equation}
to turn equation \ref{3.2-1*2} into a differential equation in the full line ($-\infty, \infty$) given by
\begin{equation}
\left[ -\frac{1}{2} \frac{d^2}{dx^2} +  \frac{e^{2x}}{2 N^2} - e^x  \right] \psi = -\frac{1}{2} (L + \frac{1}{2})^2 \psi\;.
\label{3.2-2*1}
\end{equation}

It is interesting to note that equation \ref{3.2-2*1} describes a Morse potential $\frac{e^{2x}}{2 N^2} - e^x$ with eigenvalues $-\frac{1}{2} (L + \frac{1}{2})^2$.
We next find the SUSY-partner hamiltonian of equation \ref{3.2-2*1}.

In order to be able to use the results of section \ref{susyqm}, we have to chose $V_{1}(x)$ of equation \ref{3.2-2*1} such that its ground state eigenvalue is zero.
Given $N$, the $L$ values are (from page \pageref{page11}) $N-1$, $N-2$, $N-3$, \ldots, $\left| M  \right|$. Hence, the ground state eigenvalue of equation
\ref{3.2-2*1} is $-\frac{1}{2} \left( N-1+1/2  \right)^2 = -\frac{1}{2} \left( N-1/2  \right)^2$. We rewrite equation \ref{3.2-2*1} as 
\begin{equation}
H_{1} \psi = \left[ -\frac{1}{2} \frac{d^2}{dx^2} +  \frac{e^{2x}}{2 N^2} - e^x + \frac{1}{2} \left( N-1/2  \right)^2 \right] \psi =
\left[ -\frac{1}{2} (L + 1/2)^2 + \frac{1}{2} \left( N-1/2  \right)^2\right] \psi
\label{3.2-2*2}
\end{equation}
with
\begin{equation}
V_{1}(x) = \frac{e^{2x}}{2 N^2} - e^x + \frac{1}{2} \left( N-1/2  \right)^2 
\label{3.2-2*3}
\end{equation}
chosen such that the ground state eigenvalue is zero. We are now ready to get the SUSY-partner hamiltonian of equation \ref{3.2-2*2}. From equations
\ref{3.2-2*3} and \ref{1susyqm3}, we get the Ricatti equation,
\begin{equation}
\frac{e^{2x}}{2 N^2} - e^x + \frac{1}{2} \left( N-1/2  \right)^2 = \frac{1}{2} \left[  W_{1}^{2} - \frac{dW_{1}}{dx}\right]
\end{equation}
whose solution is
\begin{equation}
W_{1} = \frac{e^x}{N} + \frac{1}{2} - N
\label{3.2-3*1}
\end{equation}

From equations \ref{3.2-3*1} and \ref{6susyqm2}, we construct 
\begin{equation}
A^{-}_{1} = \frac{1}{\sqrt{2}} \left( \frac{d}{dx} + \frac{e^x}{N} + \frac{1}{2} - N  \right)\;\;\;\;\;\;\;\; \rm and \;\;\;\;\;\;\;\; \mit 
A^{+}_{\rm 1} = \frac{\rm 1}{\sqrt{\rm 2}} \left( -\frac{d}{dx} + \frac{e^x}{N} + \frac{1}{2} - N  \right)\;.
\label{3.2-3*2/3}
\end{equation}
The SUSY-partner hamiltonian of $H_{1}$ in equation \ref{3.2-2*2} is given by equations \ref{3.2-3*2/3} and \ref{2susyqm3} yielding
\begin{equation}
H_{2}  = A^{-}_{1} A^{+}_{\rm 1} =  -\frac{1}{2} \frac{d^2}{dx^2} +  \frac{e^{2x}}{2 N^2} - \left( 1 - \frac{1}{N} \right) e^x + \frac{1}{2} 
\left( N-1/2  \right)^2  \;.
\label{3.2-3*4}
\end{equation}

From the discussion in section \ref{susyqm}, we know that $H_{2}$ has the same eigenvalues as $H_{1}$ except for the ground state where $L=N-1$
and that the eigenstates $\tilde{\psi}$ of $H_{2}$ are related to that of the eigenstates of $H_{1}$ by $\tilde{\psi} \sim A^{-}_{1} \psi$. We
then write the eigenvalue equation for equation \ref{3.2-3*4} as 
\begin{equation}
\begin{array}{lll}
H_{2} \tilde{\psi} & =  & \left[-\frac{1}{2} \frac{d^2}{dx^2} +  \frac{e^{2x}}{2 N^2} - \left( 1 - \frac{1}{N} \right) e^x + \frac{1}{2} 
\left( N-1/2  \right)^2 \right] \tilde{\psi} \\
 &  = & \left[ -\frac{1}{2} (L + 1/2)^2 + \frac{1}{2} \left( N-1/2  \right)^2\right] \tilde{\psi} \;.
\label{3.2-3*5}
\end{array}
\end{equation}

Summarizing, we have the SUSY-partner eigenvalue equations given by equations \ref{3.2-2*2} and \ref{3.2-3*5} (with the rescaling constant term
$\frac{1}{2} \left( N-1/2  \right)^2$ cancelled out)
\begin{equation}
\begin{array}{lll}
{\rm a)}\; & \left[ -\frac{1}{2} \frac{d^2}{dx^2} +  \frac{e^{2x}}{2 N^2} - e^x  \right] \psi =
 -\frac{1}{2} (L + 1/2)^2  \psi \\
{\rm b)}\; & \left[-\frac{1}{2} \frac{d^2}{dx^2} +  \frac{e^{2x}}{2 N^2} - \left( 1 - \frac{1}{N} \right) e^x \right] \tilde{\psi} 
 = -\frac{1}{2} (L + 1/2)^2  \tilde{\psi}\; .
\end{array}
\label{3.2-4*1}
\end{equation}

Transforming back to the variable $r$ and the eigenstate $u_{NL}$ using equations \ref{3.2-1*1} and \ref{3.2-1*3}, we get from equation \ref{3.2-4*1}
\begin{equation}
\begin{array}{lll}
{\rm a)}\; & \left[ -\frac{1}{2} \frac{d^2}{dr^2} - \delta \frac{\mu e^2}{\hbar^2} \frac{1}{r} + \frac{L (L+1)}{2 r^2}  \right] u_{NL} =
  - \frac{\delta^2}{N^2} \frac{\mu^{2} e^4}{2 \hbar^4} u_{NL} \\
{\rm b)}\; & \left[ -\frac{1}{2} \frac{d^2}{dr^2} - \delta^{\prime} \frac{\mu e^2}{\hbar^2} \frac{1}{r} + \frac{L (L+1)}{2 r^2}  \right] \tilde{u}_{N^{\prime}L} =
  - \frac{\delta^{\prime 2}}{N^{\prime 2}} \frac{\mu^{2} e^4}{2 \hbar^4} \tilde{u}_{N^{\prime}L}
\end{array}
\label{3.2-4*2}
\end{equation}
where
\begin{equation}
\begin{array}{lll}
{\rm a)}\; & \delta \equiv \eta \sigma^2 \\
{\rm b)}\; & \delta^{\prime} = \left( 1 - \frac{1}{N} \right) \delta \\
{\rm c)}\; & N^{\prime} = N - 1
\end{array}
\label{3.2-4*3}
\end{equation}

From equations \ref{3.2-4*2} and \ref{3.2-4*3}, it becomes apparent that if we formulate the SUSYQM problem in the full line by the change of variables given by 
equation \ref{3.2-1*3}, states with the quantum numbers ($N$, $L$) in a potential with parameter $\eta \sigma^2$ is the SUSY partner of states with quantum numbers
($N-1$, $L$) but in a potential with parameter $\left( 1 - \frac{1}{N} \right) \eta \sigma^2$. This relationship is in sharp contrast to that of subsection \ref{eigen} 
in which states with quantum numbers ($N$, $L$) with a potential having parameter $\eta \sigma^2$  has as their SUSY partners, states with quantum numbers ($N$, $L+1$)
with the potential having the \underline{same} parameter $\eta \sigma^2$. By SUSYQM, we have related states with different $\eta \sigma^2$ and $N$ as SUSY partners in the Hartmann 
potential.

An illustration of the observations put forth in the preceding paragraph is illustrated in figure \ref{fig6eta*3.2-5} for $N = \left| M \right| + 3$ in which  
$L = \left| M \right| + 2,\; \left| M \right| + 1,\; \left| M \right| $. The SUSY-partner states will have $N^{\prime} = \left| M \right| + 2$ with 
$L =  \left| M \right| + 1,\; \left| M \right| $. If $\eta \sigma^2$ is the value for the first set of eigenstates, then its SUSY partners will have a value of 
$\left( 1- \frac{1}{N} \right) \eta \sigma^2 = \left( \frac{\left| M \right| + 2}{\left| M \right| + 3}  \right) \eta \sigma^2$. Note that the actual SUSY eigenvalues
are given by $-\frac{1}{2} \left( L + \frac{1}{2}  \right)^2$ as in equation \ref{3.2-4*1}. In addition, 
note carefully that in equation \ref{3.2-4*2}, the \underline{energy} 
eigenvalues of the SUSY-partner eigenvalue equations are \underline{identical} since from equation \ref{3.2-4*3} 
\begin{equation}
\frac{ \delta^{\prime} }{ N^{\prime}} = \frac{\left( 1 - \frac{1}{N} \right) \delta}{N-1} = 
\frac{\left( N-1 \right) \delta /N}{N-1} = \frac{\delta}{N}\;.
\end{equation}
Hence, in figure \ref{fig6eta*3.2-5}, the eigenstates have the same energy eigenvalue but different SUSY eigenvalues. We have here a case in which the SUSY partnership does not
involve the actual energy eigenvalues.

Another thing to note from figure \ref{fig6eta*3.2-5} is that indeed, the spectrum of states of the Fermi sector has all the corresponding states of the Bosonic sector except for its
ground state, as expected.
\clearpage

\section{Conclusion}
\label{concl}
\indent

In the preceding discussions, we have demonstrated how SUSYQM techniques can be used in the radial equation of the Hartmann potential in theoretical
chemistry. By formulating SUSYQM in the half line $\left[ 0,\infty \right)$, we are able to establish a connection between the states with quantum numbers
($N$, $L$) and ($N$, $L+1$) and the same parameter values $\eta \sigma^2$. This enabled us to derive the energy eigenvalues and radial eigenfunctions.
On the other hand, formulating SUSYQM in the full line $\left( -\infty,\infty \right)$, established an interesting SUSY connection between states with 
quantum numbers ($N$, $L$) and the parameter value $\eta \sigma^2$ with that of states with quantum numbers ($N-1$, $L$) and the parameter value 
$\left( 1 - \frac{1}{N} \right) \eta \sigma^2$.

The first formulation basically tells us that SUSYQM techniques can be used as an alternative method of solving the Schr\"odinger equation. The second 
formulation reveals the possibility of unraveling new and unexpected relationships between eigenstates with different parameters and quantum numbers.

The key result in SUSYQM is the intimate relationship of the eigenvalues and eigenfunctions of the hierarchy of SUSY-partner hamiltonians. This can be very 
useful in solving the Schr\"odinger equation of a complicated hamiltonian if its SUSY-partner hamiltonian is easily solvable.

A very useful result in the present discussion is the fact that $A^{-} \psi^{0} = 0$. This enabled us to solve a first order differential
equation (as in equation \ref{45p}) instead of the second order Schr\"{o}dinger differential equation to give us the eigenfunctions and eigenvalues of the states 
at the lowest rung of the tower of states of each of the hamiltonians in the hierarchy. The rest of the eigenfunctions and eigenvalues 
are then easily computed by applying the corresponding $A^{+}_{L}$ operators to these eigenfunctions.

As already pointed out in this article, the separated equations of the Hartmann potential and the hydrogen atom greatly resemble each other. 
A number of studies of the SUSY features of the 
Coulomb problem in hydrogenic atoms have been made over the past years \cite{bluhm,kost,nieto}. These studies may very well lead to
some further insights into the workings of SUSY in the Hartmann potential due to the similarity of its separated equations with that of
the hydrogen atom.

With the above considerations, the author hopes to stimulate further examples of applications of SUSYQM in important problems in theoretical chemistry.
\clearpage

\section*{Figure Captions}
\begin{enumerate}
   \item The hierarchy of hamiltonians and the action of the operators $A^{\pm}_{n}$ on the degenerate eigenstates
   \item The energy states of the hierarchy of SUSY-partner hamiltonians from the Hartmann potential
   \item The hierarchy of hamiltonians of the Hartmann potential and their ground states. The $H_{L}$ here are the actual radial hamiltonian for a particular $L$ value.
   \item Figure 3 with the energy levels labelled.
   \item The energy eigenstates of the Hartmann potential. The action of the $A^{+}_{L}$ operators are explicitly shown to indicate how the other states are
obtained from the states at the lowest rung of the hierarchy of hamiltonians.
   \item An illustration of SUSY-partner eigenstates identified by ($N$, $L$, $\eta \sigma^2$) and 
         $\left( N-1,\; L,\right.$ $ \left. \left( 1-\frac{1}{N}  \right)\eta \sigma^2  \right)$
         given $N = \left| M  \right| + 3$ for the SUSYQM formulation 
        of the Hartmann potential in the full line ($-\infty$,$\infty$). Of course, one can again relate the
        different eigenstates with the same SUSY eigenvalues by $A^{\pm}_{1}$ of equation \ref{3.2-3*2/3}.
\end{enumerate}
\clearpage

%%%%%%%%%%%%%%%%%%%%%%%%%%%%%%%%%%%%%%%%%%%%%%%%%%%%%%%%%%%%%%%%%%%%%%%%%%%%%%%%%%%%%%%%%%%%%%%%%%%%%%%%%%%%%%%%%%%%%%%%%%%%%%%%%%%%%%
\clearpage
%%%%%%%%%%%%%%%%%%%%%%%%%%%%%%%%%%%%%%%%%%%%%%%%%%%%%%%%%%%%%%%%%%%%%%%%%%%%%%%%%%%%%%%%%%%%%%%%%%%%%%%%%%%%%%%%%%%%%%%%%%%%%%%%%%%%%%

%%%%%%%%%%%%%%%%%%%%%%%%%%%%%%%%%%%%%%%%%%%%%%%%%%%%%%%%%%%%%%%%%%%%%%%%%%%%%%%%%%%%%%%%%%%%%%%%%%%%%%%%%%%%%%%%%%%%%%%%%%%%%%%%%%%%%%
\clearpage
%%%%%%%%%%%%%%%%%%%%%%%%%%%%%%%%%%%%%%%%%%%%%%%%%%%%%%%%%%%%%%%%%%%%%%%%%%%%%%%%%%%%%%%%%%%%%%%%%%%%%%%%%%%%%%%%%%%%%%%%%%%%%%%%%%%%%%

%%%%%%%%%%%%%%%%%%%%%%%%%%%%%%%%%%%%%%%%%%%%%%%%%%%%%%%%%%%%%%%%%%%%%%%%%%%%%%%%%%%%%%%%%%%%%%%%%%%%%%%%%%%%%%%%%%%%%%%%%%%%%%%%%%%%%%
\clearpage
%%%%%%%%%%%%%%%%%%%%%%%%%%%%%%%%%%%%%%%%%%%%%%%%%%%%%%%%%%%%%%%%%%%%%%%%%%%%%%%%%%%%%%%%%%%%%%%%%%%%%%%%%%%%%%%%%%%%%%%%%%%%%%%%%%%%%%

%%%%%%%%%%%%%%%%%%%%%%%%%%%%%%%%%%%%%%%%%%%%%%%%%%%%%%%%%%%%%%%%%%%%%%%%%%%%%%%%%%%%%%%%%%%%%%%%%%%%%%%%%%%%%%%%%%%%%%%%%%%%%%%%%%%%%%
\clearpage
%%%%%%%%%%%%%%%%%%%%%%%%%%%%%%%%%%%%%%%%%%%%%%%%%%%%%%%%%%%%%%%%%%%%%%%%%%%%%%%%%%%%%%%%%%%%%%%%%%%%%%%%%%%%%%%%%%%%%%%%%%%%%%%%%%%%%%

%%%%%%%%%%%%%%%%%%%%%%%%%%%%%%%%%%%%%%%%%%%%%%%%%%%%%%%%%%%%%%%%%%%%%%%%%%%%%%%%%%%%%%%%%%%%%%%%%%%%%%%%%%%%%%%%%%%%%%%%%%%%%%%%%%%%%%
\begin{figure}[htbp]
\begin{center}
\begin{picture}(400,200)
%%%%%%%%%%%%%%%%%%%%%%%%%%%%%
                                                                       \put(37.5,173){\vdots}                                                                            \put(167.5,173){\vdots}                                                                                      \put(297.5,173){\vdots}
                                                                       \put(37.5,160){\vdots}                                                                            \put(167.5,160){\vdots}                                                                                      \put(297.5,160){\vdots}
                                                                                                                                 \put(95,145){$A^{+}_{\left| M \right|}$}                                                                \put(215,145){$A^{+}_{\left| M \right| + 1}$}
                                                                                                                                 \put(130,135){\vector(-1,0){55}}                                                                        \put(260,135){\vector(-1,0){55}}
                                                                       \put(15,135){$u_{\left| M \right| + 3, \left| M \right|}$}                                        \put(140,135){$u_{\left| M \right| + 3, \left| M \right| + 1}$}                                              \put(275,135){$u_{\left| M \right| + 3, \left| M \right| + 2}$}
\put(-40,130){$\frac{-\Lambda}{\left( \left| M \right| +3 \right)^2}$} \put(0,130){\line(75,0){75}}                                                                      \put(130,130){\line(75,0){75}}                                                                               \put(260,130){\line(75,0){75}}
                                                                                                                                 \put(95,90){$A^{+}_{\left| M \right|}$}
                                                                                                                                 \put(130,80){\vector(-1,0){55}}
                                                                       \put(10,80){$u_{\left| M \right| + 2, \left| M \right|}$}                                         \put(140,80){$u_{\left| M \right| + 2, \left| M \right| + 1}$}
\put(-40,75){$\frac{-\Lambda}{\left( \left| M \right| +2 \right)^2}$}  \put(0,75){\line(75,0){75}}                                                                       \put (130,75){\line(75,0){75}}
                                                                       \put(10,25){$u_{\left| M \right| + 1, \left| M \right|}$}
\put(-40,20){$\frac{-\Lambda}{\left( \left| M \right| +1 \right)^2}$}  \put(0,20){\line(75,0){75}}
                                                                       \put(30,0){$H_{\left| M \right|}$}                                                                \put(155,0){$H_{\left| M\right| + 1}$}                                                                      \put(280,0){$ H_{\left| M \right| +  2}$}  \put(410,0){\ldots \ldots}
%%%%%%%%%%%%%%%%%%%%%%%%%%%%
\end{picture}
\end{center}
\caption{The energy eigenstates of the Hartmann potential. The action of the $A^{+}_{L}$ operators are explicitly shown to indicate how the other states are
obtained from the states at the lowest rung of the hierarchy of hamiltonians.}
\label{fig5susyhh7}
\end{figure}
\clearpage
%%%%%%%%%%%%%%%%%%%%%%%%%%%%%%%%%%%%%%%%%%%%%%%%%%%%%%%%%%%%%%%%%%%%%%%%%%%%%%%%%%%%%%%%%%%%%%%%%%%%%%%%%%%%%%%%%%%%%%%%%%%%%%%%%%%%%%

%%%%%%%%%%%%%%%%%%%%%%%%%%%%%%%%%%%%%%%%%%%%%%%%%%%%%%%%%%%%%%%%%%%%%%%%%%%%%%%%%%%%%%%%%%%%%%%%%%%%%%%%%%%%%%%%%%%%%%%%%%%%%%%%%%%%%%
\begin{figure}[htbp]
\begin{center}
\begin{picture}(400,200)
%%%%%%%%%%%%%%%%%%%%%%%%%%%%%
                            \put(0,173){SUSY eigenvalue}                                                 \put(155,173){$\eta \sigma^2$}                                  \put(270,173){$\left( \frac{\left| M \right| + 2}{\left| M \right| + 3}  \right) \eta \sigma^2$}
                                                                                                         \put(145,135){$u_{\left|M\right|+ 3,\left| M \right|}$}         \put(275,135){$\tilde{u}_{\left| M \right| + 2,\left| M \right|}$}
                            \put(0,130){$-\frac{1}{2} \left[ \left| M \right| + \frac{1}{2} \right]^2$}  \put(130,130){\line(75,0){75}}                                  \put(260,130){\line(75,0){75}}
                                                                                                         \put(140,80){$u_{\left| M \right| + 3,\left| M \right| + 1}$}   \put(270,80){$\tilde{u}_{\left| M \right| + 2, \left| M \right| + 1}$}
                            \put(0,75){$-\frac{1}{2} \left[ \left| M \right| + \frac{3}{2} \right]^2$}   \put (130,75){\line(75,0){75}}                                  \put (260,75){\line(75,0){75}}
                                                                                                         \put(140,25){$u_{\left| M \right| + 3,\left| M \right| + 2}$}
                            \put(0,20){$-\frac{1}{2} \left[ \left| M \right| + \frac{5}{2} \right]^2$}   \put(130,20){\line(75,0){75}}        
%%%%%%%%%%%%%%%%%%%%%%%%%%%%
\end{picture}
\end{center}
\caption{An illustration of SUSY-partner eigenstates identified by ($N$, $L$, $\eta \sigma^2$) and 
$\left( N-1,\; L,\; \left( 1-\frac{1}{N}  \right)\eta \sigma^2  \right)$
 given $N = \left| M  \right| + 3$ for the SUSYQM formulation 
of the Hartmann potential in the full line ($-\infty$,$\infty$) Of course, one can again relate the
        different eigenstates with the same SUSY eigenvalues by $A^{\pm}_{1}$ of equation 72.}
\label{fig6eta*3.2-5}
\end{figure}
\clearpage
%%%%%%%%%%%%%%%%%%%%%%%%%%%%%%%%%%%%%%%%%%%%%%%%%%%%%%%%%%%%%%%%%%%%%%%%%%%%%%%%%%%%%%%%%%%%%%%%%%%%%%%%%%%%%%%%%%%%%%%%%%%%%%%%%%%%%%

\clearpage
\label{ref}

\end{document}